\documentstyle[sprocl,epsf]{article}

\def\Journal#1#2#3#4{{#1} {\bf #2}, #3 (#4)}

\def\PLB{{\em Phys.\ Lett.}\ B}

\def\PRD{{\em Phys.\ Rev.}\ D}

\def\w{{\rm w}}
\def\tr{{\rm tr}}
\def\barrier{{\rm barrier}}
\def\eff{{\rm eff}}
\def\L{{\rm L}}
\def\T{{\rm T}}
\def\Im{{\rm Im}}

\def\gtrsim{\mathrel{\mathpalette\vereq>}}
\makeatletter
\def\vereq#1#2{\lower3pt\vbox{\baselineskip1.5pt \lineskip1.5pt
  \ialign{$\m@th#1\hfill##\hfil$\crcr#2\crcr\sim\crcr}}}
\makeatother

\def\ggquestion{{\buildrel ? \over \gg}}

\begin{document}

\title{B VIOLATION IN THE HOT STANDARD MODEL}

\author{ PETER ARNOLD }

\address{Department of Physics, University of Washington,
Seattle, Washington 98195
\footnote{
   Address after Sept.\ 1, 1997: Department of Physics, University of
   Virginia, Charlottesville, Virginia 22901.
}
}

\maketitle\abstracts{
I explain, as simply and pedagogically as I can, recent arguments that
the high-temperature baryon number violation rate depends on the
electroweak coupling as $\Gamma = \# \alpha_\w^5 T^4$
(up to higher order corrections).
This is
in contrast to the form $\Gamma = \# \alpha_\w^4 T^4$ that has
historically been assumed in discussions of electroweak baryogenesis.
I will give a general analysis of the time scale for
non-perturbative gauge field fluctuations in hot non-Abelian plasmas.
}


\section{Introduction}

I will be discussing baryon number (B) violation in the Standard Model in
the hot, early universe.  The motivation is scenarios of electroweak
baryogenesis, which typically depend on the rate $\Gamma$ of B violation
in the hot, symmetric phase of electroweak theory.  The particular focus
of my discussion will be to understand what physics determines the
size of $\Gamma$ and to understand how $\Gamma$ depends
on the parameters of the theory.

The source of B violation
is the electroweak anomaly for baryon number, which demands that
\begin {equation}
   \Delta B \sim g_\w^2 \int dt \; d^3x \> \tr F_\w \tilde F_\w
   .
\label {eq:anomaly}
\end {equation}
The subscript ``w'' (often suppressed in the following)
indicates that the
coupling is the weak coupling and the field strengths those of the
weak gauge fields.
This anomaly equation relates the amount of B violation to
topological transitions in the weak gauge fields.
I shan't review the anomaly any further but will just take
(\ref{eq:anomaly}) as my starting point.

The problem of understanding the size of the B violation rate can be
simplified a bit.
Because the physics of interest is the B violation rate in the hot,
{\it symmetric} phase of electroweak theory, the Higgs field doesn't
play any essential role.  So I shall ignore the Higgs field altogether.
In addition, even though the behavior of the fermions is the ultimate
matter of interest, fermions don't actually play an essential role
in determining the rate.  The anomaly (\ref{eq:anomaly}) tells us
that fermion number violation is a consequence of what the {\it gauge}
fields are doing: If the gauge fields have non-trivial
$g^2 \int F \tilde F$, then B violation will tag along as a consequence.
As long as couplings are small, the back-reaction of the fermions
on the gauge-field dynamics turns out to be a higher-order effect.
So one can forget about the fermions and analyze the following, simplified
problem:~%
\footnote{
   It is possible to convert the rate in pure
   SU$_2$ gauge theory into
   the rate in the real theory that includes the Higgs and fermions.
   See sec.\ II of ref.\ 1
   for the conversion.
}
\begin {quote}
   {\it
   What's the rate of topological transitions in \underline{pure}
   SU$_2$ gauge theory at high temperature?
   }
\end {quote}

By dimensional analysis, the rate per unit volume at high temperature must
depend on temperature as $T^4$.  But what is the dependence on $\alpha_\w$?
The purpose of this talk is to explain a relatively recent analysis by
myself, Dam Son, and Larry Yaffe\,\cite{alpha5}
that, in weak coupling ($\alpha_\w\to0$),
the B violation rate behaves as
\begin {equation}
   \Gamma \equiv {\hbox{rate}\over\hbox{volume}}
   = \# \alpha_\w^5 T^4
   .
\end {equation}
In contrast, it has instead long been assumed in the literature that
$\Gamma = \# \alpha_\w^4 T^4$.  I shall explain later the origin of
this belief.


\section{The scales of the problem}

  Because of the factor of $g^2$ in the anomaly equation (\ref{eq:anomaly}),
it follows that $\Delta B$ of order 1 requires non-perturbatively large
field strengths of order $1/g$.  The energy of the intermediate gauge
configurations in a B-violating process will therefore also be 
non-perturbatively large and of order $1/g^2$.  The moral to keep
in mind in everything that follows is that
{\it non-perturbatively large} amplitude fluctuations are required for
a topological transition.

\begin {figure}
\vbox
   {%
   \begin {center}
      \leavevmode
      \def\epsfsize #1#2{0.50#1}
      \epsfbox [150 260 500 530] {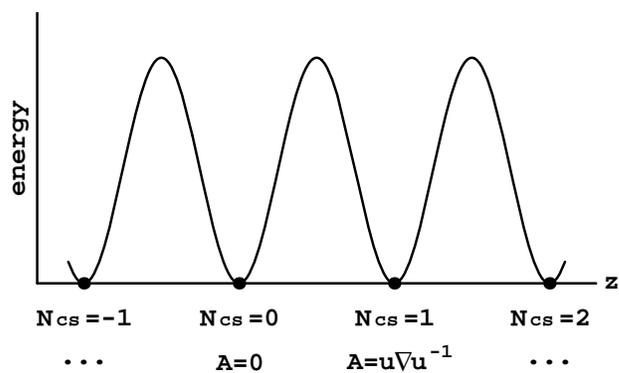}
   \end {center}
   \caption
       {%
       A schematic representation of
       the (bosonic) potential energy along a particular direction (labeled
       $z$) in field space, corresponding to topologically non-trivial
       transitions
       between vacua.
       $N_{\rm cs}$ labels the Chern-Simmons number corresponding to the
       different large gauge transformations of the $\vec A = 0$ vacuum.
       \label{fig:ridge1}
       }%
   }%
\end {figure}

  Now consider fig.~\ref{fig:ridge1},
which is a very rough depiction of the
potential energy in gauge configuration space (and is required by
international law to be shown in every talk on electroweak B violation).
The horizontal axis depicts one particular direction in the
infinite-dimensional space of gauge field configurations $\vec A(x)$.
(I'm working for the moment in $A_0=0$ gauge.)  The vertical axis
corresponds to the potential energy of those configurations.
The minima correspond to the perturbative vacuum $\vec A = 0$ and
large gauge transformations of it, and you can think of the configurations
inbetween as those that a particular process passes through when
violating B through the anomaly.  There is therefore an energy barrier
$E_\barrier$ of order $1/g^2$ for such processes.  At non-zero temperature, one
expects that the probability for the system to have enough energy to
pass over this energy barrier is given roughly by a Maxwell-Boltzmann factor,
\begin {equation}
   \Gamma \sim \exp\left(-\beta E_\barrier\right)
   ,
\end {equation}
where $\beta$ is the inverse temperature.

\begin {figure}
\vbox
   {%
   \begin {center}
        \leavevmode
        \def\epsfsize #1#2{0.51#1}
        \epsfbox [150 200 500 500] {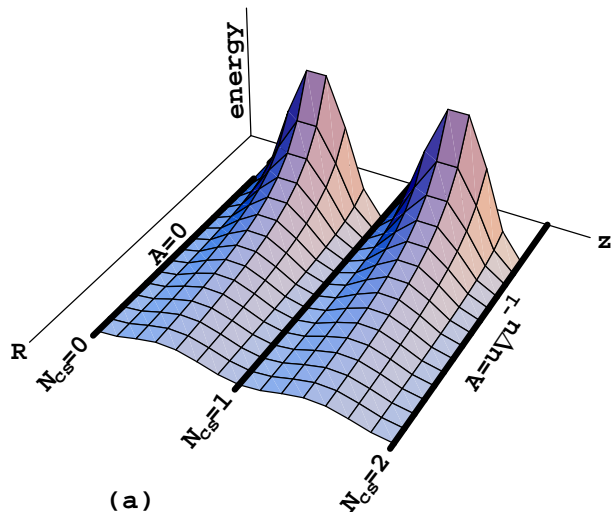}
   \end {center}
   \caption
       {%
       The same as fig.~\protect\ref{fig:ridge1} but supplemented by an
       extra dimension of configuration space corresponding to
       the spatial size $R$ of the configurations.
       \label{fig:ridge2}
       }%
   }%
\end {figure}

  So far I've discussed the dependence of the fields and energy barrier on
coupling $g$, but I haven't explained what sets the dimensionful scale
for these quantities.  Pure gauge theory is (at least classically) scale
invariant.  If the gauge configurations responsible for a topological
transition have spatial size $L$, then $L$ is the only thing that can
set the scales.  $\Delta B \sim 1$ then implies
\begin {equation}
   F_\barrier \sim {1\over g L^2} ,
   \qquad
   A_\barrier \sim {1\over g L} ,
   \qquad
   E_\barrier \sim {1\over g^2 L} .
\label {eq:L}
\end {equation}
To capture the most important qualitative features of the gauge configuration
potential energy, we should really supplement Fig.~\ref{fig:ridge1}
by one additional
direction in gauge configuration space, representing the spatial size
$L$ of configurations.
This gives Fig.~\ref{fig:ridge2}, where the energy barrier has
now become a ridge.
As given by (\ref{eq:L}), transitions through very small configurations are
associated with a high energy barrier, while transitions through
arbitrarily large configurations are associated with an arbitrarily small
barrier.
The transition probability
\begin {equation}
   \Gamma \sim \exp(-\beta E_\barrier) \sim \exp(-\#/g^2 L T)
\end {equation}
is unsuppressed when $L \gtrsim 1/g^2 T$.  In fact, $L \sim 1/g^2 T$ is
the dominant scale associated with transitions.  This is due to
entropy: there are a lot more short-distance modes than long-distance
modes.  As a result, transitions are dominated by the shortest
distance scale $L$ that is unsuppressed.
\begin {equation}
\matrix{
   \hbox{Dominant spatial scale:} \hfill &  L \sim 1/g^2 T\hfill  \cr
   \hbox{Dominant momentum scale:}\hfill &  k \sim g^2 T  \hfill  \cr
}
\end {equation}
This is, more generally, the scale for {\it any} unsuppressed,
non-perturbative, colored fluctuations.

We can now understand the naive rate estimate that has previously
been used in the literature.  In a relativistic theory of massless
particles, one might expect time scales to be the same as distance
scales, so that
\begin {equation}
   t \sim L \sim {1\over g^2 T}
   \,.
   \qquad\qquad
   \hbox{~~~~(naive)}
\end {equation}
A rate of one unsuppressed transition per volume $L^3$ per time $t$
then gives
\begin {equation}
   \Gamma \sim {1\over L^3 t} \sim g^8 T^4
   .
   \qquad\qquad
   \hbox{(naive)}
\end {equation}
This would predict $\Gamma = \# \alpha^4 T^4$ where $\#$ is a
constant that cannot be computed in perturbation theory (since the
gauge fluctuations are non-perturbative).

The problem with the above estimate is that finite temperature physics
is not Lorentz invariant.  There is a preferred frame: the rest frame
of the plasma.  As a result, one cannot assume that time scales will
be the same as distance scales.  In particular, Son, Yaffe and I have
argued that viscous forces in the plasma slow down the dynamics of the
long-distance ($L \sim 1/g^2 T$) modes.  As I shall discuss, we find that
the time scale is slowed down to $t \sim 1/g^4 T$.
\begin {equation}
\matrix{
   \hbox{Dominant time scale:}     \hfill &  t \sim 1/g^4 T\hfill  \cr
   \hbox{Dominant frequency scale:}\hfill &  \omega \sim g^4 T  \hfill  \cr
}
\end {equation}
As a result, the rate $\Gamma \sim 1/t L^3$ is then
\begin {equation}
   \Gamma = \# \alpha^5 T^4 \,,
\end {equation}
where $\#$ is again a non-perturbative constant.


\begin {figure}
\vbox
   {%
   \begin {center}
      \leavevmode
      \def\epsfsize #1#2{0.50#1}
      \epsfbox {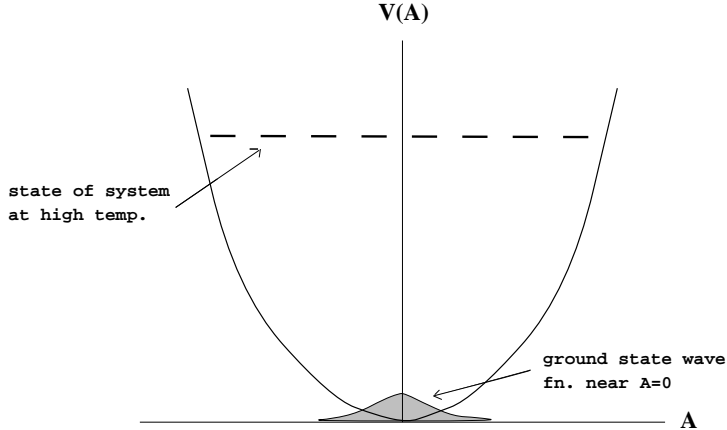}
   \end {center}
   \caption
       {%
       A slightly anharmonic oscillator.  High-energy states probe
       non-perturbatively large amplitudes.
       \label{fig:highT}
       }%
   }%
\end {figure}

\section {Why high T generically creates non-perturbatively large
amplitude fluctuations}

I'd now like to explain very generally why physics becomes
non-perturbative at high temperature, even in theories that are
weakly coupled ($g \ll 1$).  As a simple example, consider a very
simple problem from quantum or classical mechanics: a very slightly
anharmonic oscillator.  The Lagrangian is simply
\begin {equation}
   L \sim \dot \phi^2 - V(\phi) ,
   \qquad\qquad
   V(\phi) \sim m^2 \phi^2 + g^2 \phi^4 .
\end {equation}
At zero temperature and low energies, the states of interest are the
ground state wave function and its low-lying excitations.  The wave
function is localized near $\phi=0$, and the anharmonic term in the
potential can be treated as a perturbation.  At arbitrarily high
temperature, however, the system occupies states of arbitrarily high energy,
which are states that probe arbitrarily large amplitudes of $\phi$.
But, for large enough $\phi$, a quartic term $g^2 \phi^4$ will always
dominate over a quadratic one $m^2 \phi^2$, no matter how small $g$ is.
The situation is depicted schematically in Fig.~\ref{fig:highT}.

To estimate when perturbation theory breaks down, first pretend the
anharmonic term can indeed be treated perturbatively.  For a harmonic
oscillator, the amplitude of $\phi$ at high temperature is given by
the equipartition theorem:
\begin {equation}
   m^2 \phi^2 \sim T 
   \qquad\qquad\Longrightarrow\qquad\qquad
   \phi^2 \sim {T\over m^2} .
\label {eq:phi estimate}
\end {equation}
A measure of whether $\phi$ is perturbative is given by the ratio of
the anharmonic to harmonic terms:
\begin {equation}
   1 \ggquestion {g^2 \phi^4 \over m^2 \phi^2} \sim {g^2 T \over m^4} \,.
\label {eq:pert condition}
\end {equation}
The thing to keep in mind is that large $T$, or equivalently small $m$,
implies large, non-perturbative amplitudes $\phi$.  Conversely, small
$T$ or large $m$ implies small, perturbative amplitude fluctuations in
$\phi$.

The field theory versions of (\ref{eq:phi estimate}) and
(\ref{eq:pert condition}) look quite similar, with the natural frequency
$m$ of the oscillator
replaced by $\omega_k$, where $k$ is the spatial momentum characteristic
of the modes of interest.  Specifically, the amplitude analogous to
(\ref{eq:phi estimate}) is
\begin {equation}
   \left[ (\bar\phi)_L \right]^2 \sim {T \over \omega_k^2 L^3}
\end {equation}
where $(\bar\phi)_L$ represent $\phi$ smeared (i.e. averaged) over a
spatial scale of order $L$ and where
\begin {equation}
   k \sim 1/L .
\end {equation}
The perturbative condition analogous to (\ref{eq:pert condition}) is
\begin {equation}
   1 \ggquestion {g^2 T \over \omega_k^4 L^3}
   .
\end {equation}
The amplitude of fluctuations will be perturbative on {\it every} distance
scale $L$ if
\begin {equation}
   1 \ggquestion \max_L \left( g^2 T \over \omega_k^4 L^3 \right)
   \sim {g^2 T \over m} .
\label{eq:FT pert condition}
\end {equation}
People familiar with high-temperature perturbation theory will recognize
this as the standard condition for the ultimate success of
perturbation theory.


\begin {figure}
\vbox
   {%
   \begin {center}
      \leavevmode
      \def\epsfsize #1#2{0.50#1}
      \epsfbox {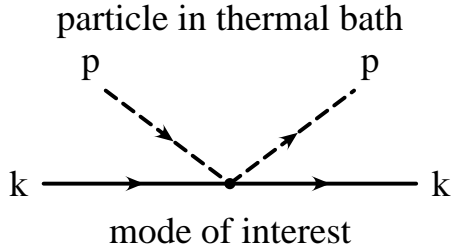}
   \end {center}
   \caption
       {%
       Forward scattering of a mode of interest (solid line) off of
       a particle in the plasma (dashed line) in $\phi^4$ theory.
       \label{fig:forward1}
       }%
   }%
\end {figure}

\section {The last section is a lie}

The last section suggested that non-perturbative amplitudes are generic
at very high temperature.  In fact, in field theories, the opposite is
true: amplitudes generically remain perturbative.
The reason is that interactions
in field theory can cause the effective mass of modes to change at high
temperature:
\begin {equation}
   m \qquad \longrightarrow \qquad m_\eff(T)
   .
\end {equation}
This change in the effective mass is caused by the forward scattering
of propagating modes off of particles in the thermal bath.  This forward
scattering modifies the effective propagation of the modes in a way that
can change their effective mass.  For definiteness, consider $\phi^4$
theory.  The leading-order forward-scattering amplitude is shown in
Fig.~\ref{fig:forward1},
where the solid line denotes the mode of interest and the
dashed line represents a particle present in the thermal bath.  As far
as the mode of interest is concerned, such a process looks just like a
mass insertion, and it turns out to give
\begin {equation}
   m_\eff^2 \sim g^2 T^2
   .
\end {equation}
But this effective mass is large enough at high $T$ that $\phi$ remains
perturbative!  The perturbative expansion parameter
(\ref{eq:FT pert condition}) becomes
\begin {equation}
   {g^2 T \over m_\eff} \sim g
   ,
\end {equation}
which is indeed small if the theory is weakly coupled.


\begin {figure}
\vbox
   {%
   \begin {center}
      \leavevmode
      \def\epsfsize #1#2{0.60#1}
      \epsfbox {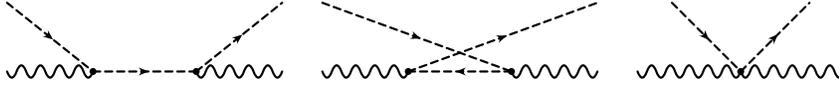}
   \end {center}
   \caption
       {%
       Forward scattering of a mode of interest (wavy line) off of
       a particle in the plasma (dashed line) in gauge theory.
       the particle in the plasma might be anything---a gauge boson,
       scalar, or fermion.
       \label{fig:forward2}
       }%
   }%
\end {figure}

\section {Gauge theory}

Now let's consider gauge theory.  The leading-order processes for
forward scattering are shown in Fig.~\ref{fig:forward2}, where the dashed line
represents any sort of particle in the thermal bath---gauge, scalar,
or fermion---that the gauge mode of interest scatters off of.
We'd like to know what sort of effective thermal mass, or more generally
what sort of self-energy $\Pi(\omega,k)$, is generated by this
forward scattering.  To physically motivate the result, it will be
useful to first review some basic features of plasmas.  Think back
to your graduate electrodynamics course and a book like Jackson.

\begin{itemize}

\item
{\it Debye screening of electric fields.}
Static electric fields are screened in a plasma because the charged
particle in the plasma re-arrange themselves.
In the case at hand, this manifests as
\begin {equation}
   \Pi_\L(\omega,k) = O(g^2 T^2)
   \qquad\hbox{for}\qquad
   \omega = 0
   \qquad\qquad
   (k \ll T)
   .
\label {eq:debye}
\end {equation}
That is, the self-energy has an effective mass in the static limit
$(\omega=0)$.  The subscript $\L$ denotes the ``longitudinal'' polarization
of the gauge field, which is the polarization corresponding to electric
fields in the static limit
(polarization $A_0$ for $\omega=0$ in covariant gauge).
The $T^2$ in (\ref{eq:debye}) comes from dimensional analysis in the
ultrarelativistic limit, and the $g^2$ comes from the interactions
in fig.~\ref{fig:forward2}.

\item
{\it No screening of static magnetic fields.}
Unlike electric fields, static magnetic fields are not screened in
a plasma, and
\begin {equation}
   \Pi_\T(\omega,k) = 0
   \qquad\hbox{for}\qquad
   \omega = 0
   \qquad\qquad
   (k \ll T)
   .
\end {equation}
There is no effective mass.  The subscript $\T$ denotes the two ``transverse''
polarizations---those which correspond to magnetic fields in the
static limit.  Sometimes, people speak loosely of a ``magnetic screening
mass'' in hot non-Abelian gauge theories.  This is somewhat misleading
language, however: magnetic fields actually
become strong and confining at large
distances.%
\footnote{
   In particular, large spatial Wilson loops have area law behavior at
   high temperature, which means that magnetic interactions between
   fundamental-charge particles are strong at large distances, just like
   in zero-temperature QCD.
   Magnetic fields can only be called screened in the same sense
   that, in zero temperature QCD, gluons are ``screened'' by confinement.
}

\item
{\it Plasma frequency for propagating waves.}
In a QED plasma, propagating electromagnetic waves have a minimum
frequency.  This frequency gap is known as the plasma frequency
(or the plasmon mass).  In our case, the plasma frequency, and
in fact the frequency of all
low-momentum $(k \ll T)$, propagating, colored modes,
is of order
\begin {equation}
   \Pi_{\L,\T}(\omega,k) = O(g^2 T^2)
   \qquad\hbox{for}\qquad
   \omega > k
   \qquad\qquad
   (\omega,k \ll T)
   .
\end {equation}
 
\end {itemize}

There is an exact formula\,\cite{pi}
for the one-loop self-energy of Fig.~\ref{fig:forward2},
the details of which won't be necessary for this discussion.
A simple summary of the relevant features, which incorporate the
various limits discussed above, is
\begin {eqnarray}
   \Pi_\L &=&
   g^2 T^2
   \cases{ O(1) - i O\left(\omega\over k\right) , & ~~~~$\omega \ll k$ , \cr
           O(1)                                 , & ~~~~$\omega \ge k$ ; \cr
   }
\nonumber\\
   \Pi_\T &=&
   g^2 T^2
   \cases{ O\left(\omega^2\over k^2\right)
                - i O\left(\omega\over k\right) , & $\omega \ll k$ , \cr
           O(1)                                 , & $\omega \ge k$ . \cr
   }
\label {eq:pi}
\end {eqnarray}
Note that $\Pi$ has an imaginary part\,%
\footnote{
   There are also imaginary parts in other kinematic regimes (namely
   the $1{\to}2$ particle cut) that are
   sub-leading in temperature and which have not been shown.
}
for $\omega < k$, which I'll
explain below, and that this imaginary part dominates $\Pi_\T$ when
$\omega \ll k$.

We're now in a position to see the problem with the original assumption
in the literature that the time scale for topological transitions is
$t \sim 1/g^2 T$ so that the rate is $\Gamma \sim \alpha^4 T^4$.
The naive assumption $t \sim L$ is $\omega \sim k$ in frequency space.
But then (\ref{eq:pi}) gives $\Pi \sim g^2 T^2$, which is basically
determined by having to reproduce the plasma frequency effect.
However, $\Pi \sim g^2 T^2$ is the same as the scalar $\phi^4$ theory
case discussed in the last section, which we learned only produces
perturbatively small fluctuations in field amplitude.  So $t \sim 1/g^2 T$
cannot be the time scale for the non-perturbative gauge-field fluctuations
required for a topological transition.

\begin {figure}
\vbox
   {%
   \begin {center}
      \leavevmode
      \def\epsfsize #1#2{0.40#1}
      \epsfbox [150 260 500 530] {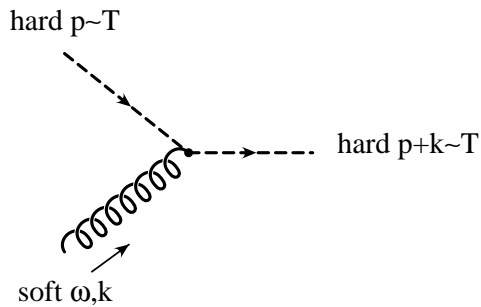}
   \end {center}
   \caption
       {%
         The absorption of a soft excitation (wavy line) by the small-angle
         scattering of a hard excitation (dashed line).
       \label{fig:absorb}
       }%
   }%
\end {figure}

Our only hope is to make the effective mass of fluctuations substantially
smaller than $\Pi \sim g^2 T^2$.  Now remember that $\Pi$ vanishes for
static magnetic fields; so it can be made arbitrarily small by considering
{\it nearly static} magnetic fields.  Eq.\ (\ref{eq:pi}) gives
\begin {equation}
   \Pi_\T \sim -i g^2 T^2 {\omega\over k}
   \qquad\qquad
   (\omega \ll k)
   .
\end {equation}
This can also be written as
\begin {equation}
   \Pi_\T \sim {g^2 T^2\over k} \, \partial_t
   .
\end {equation}
This appears as a damping term in the effective equations of
motion for the gauge field.  The physical origin of this damping
is the process shown in fig.~\ref{fig:absorb}.  Low-momentum, space-like quanta
corresponding to nearly-static magnetic fields can be absorbed by
particles in the plasma, with a resulting loss of energy from
the magnetic fields of interest.

Most of the particles in the plasma have large momenta $p\sim T \gg g^2 T$.
The damping effect just discussed for low-momentum magnetic fields is indeed
dominated by the contribution from absorption by high-momentum particles.


\begin {figure}
\vbox
   {%
   \begin {center}
      \leavevmode
      \def\epsfsize #1#2{0.50#1}
      \epsfbox [0 260 500 650] {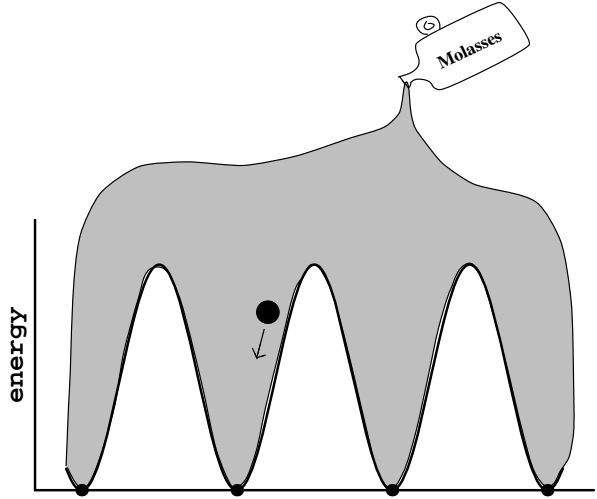}
   \end {center}
   \caption
       {%
       Similar to fig.\ \protect{\ref{fig:ridge1}} except now the dynamics
       of the long-distance modes of interest is slowed down by hot molasses.
       \label{fig:molasses}
       }%
   }%
\end {figure}

\section{A non-linear pendulum in hot molasses}

As just discussed, the short-distance modes ($p\sim T$)
in the plasma cause damping of
the dynamics of the long-distance modes ($k\sim g^2 T$)
of interest for topological
transition.  A fairly good analogy is that the long-distance modes are
like a ball rolling around in a potential that has been filled with hot
molasses, as depicted in fig.~\ref{fig:molasses},
where the molasses represents the
effect of the short-distance modes.  In fact, this analogy is useful
enough that it is worth forgetting about gauge theory for a moment and
reviewing the physics of a slightly non-linear pendulum in hot molasses.

Consider the motion of a slightly non-linear pendulum---that is, the
behavior near the bottom of one of the wells in fig.~\ref{fig:molasses}:
\begin {equation}
   (\partial_t^2 + k^2)A + O(A^2) = 0
   .
\end {equation}
For a pendulum in (idealized) hot molasses, this equation should be
modified to a Langevin equation of the form
\begin {equation}
   (\partial_t^2 + \gamma \partial_t + k^2)A + O(A^2) = \xi(t)
   .
\label{eq:molasses}
\end {equation}
A damping term, with some coefficient $\gamma$, has been added on the
left side to represent the viscous drag of the molasses.
If the molasses is hot, then random thermal fluctuations of the molasses
will randomly buffet the pendulum.
This effect is represented by a random force term $\xi(t)$ on the
right-hand side.

Non-perturbative physics corresponds to large-amplitude fluctuations of
the pendulum, so that the non-linear terms in (\ref{eq:molasses}) become
important.  Let's estimate the time scales involved by first ignoring
the non-linearities and finding the time scale for the largest amplitude
fluctuations of a simple harmonic oscillator in hot molasses.

The damping of the pendulum and the random force term are both due to
interactions with the molasses, and so their strength is related.
More precisely, the two are related by the fluctuation-dissipation
theorem, whose consequence in this case (ignoring non-linear interactions)
is that the random force has a white-noise spectrum normalized to
\begin {equation}
   \langle \xi(t) \xi(t') \rangle = 2 \gamma T \delta(t-t') \,.
\label{eq:random}
\end {equation}
Note that the magnitude is proportional to both the damping
and the temperature.  The response of the oscillator (\ref{eq:molasses})
to this force is easy to solve by Fourier transform:
\begin {equation}
   A(\omega) = {\xi(\omega) \over -\omega^2 - i\gamma\omega + k^2} \,.
\end {equation}
Using (\ref{eq:random}), the power spectrum of amplitude fluctuations is then
\begin {equation}
   \langle A(\omega)^* A(\omega) \rangle =
   {2 \gamma T \, \delta(\omega-\omega') \over
      | -\omega^2 - i\gamma\omega + k^2 |^2 }
   .
\label{eq:power}
\end {equation}

\begin {figure}
\vbox
   {%
   \begin {center}
      \leavevmode
      \def\epsfsize #1#2{0.60#1}
      \epsfbox [150 300 500 530] {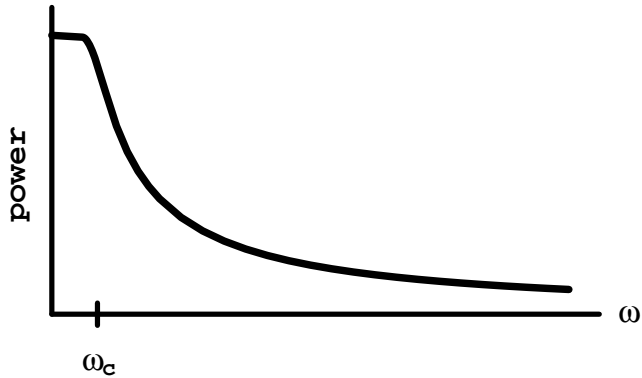}
   \end {center}
   \bigskip
   \caption
       {%
       A schematic plot of the power of amplitude fluctuations for an
       oscillator in idealized molasses.
       \label{fig:power}
       }%
   }%
\end {figure}

Now let's specialize to the case of a strongly damped pendulum, where
the damping coefficient $\gamma$ is large compared to the natural
frequency $k$.
The behavior of (\ref{eq:power}) is shown in Fig.~\ref{fig:power}:
there are small amplitude fluctuations at high frequency and larger
amplitude fluctuations at low frequency.
The largest amplitude fluctuations are when the denominator in
(\ref{eq:power}) becomes small at low frequency.
The $\omega^2$ term in the denominator can then be ignored, and the
characteristic frequency $\omega_{\rm c}$
of the largest amplitude fluctuations comes
from equating the magnitudes of the $-i\gamma\omega$ and $k^2$ terms:
\begin {equation}
   \omega_{\rm c} \sim {k^2\over\gamma} \ll k
   ,
\label{eq:wc}
\end {equation}
where the last inequality follows from the assumption $\gamma \gg k$ of
strong damping.  Note that $\omega \ll k$ is precisely the kinematic
regime of slowly varying fields that we've already discussed as being
important for non-perturbative physics in gauge theory.


\section {Back to Hot Gauge Theory}

In hot gauge theory, the effective equation for the long-distance modes
is
\begin {equation}
   (-\omega^2 + \Pi(\omega,k) + k^2) A + O(A^2) = \xi(\omega,k)
   ,
\label{eq:motion}
\end {equation}
where $\xi$ is a random force caused by interaction with the
short-distance modes.
For $\omega \ll k$, we reviewed earlier that
\begin {equation}
   \Pi(\omega,k) \approx -i \# g^2 T^2 {\omega\over k} \equiv \gamma
   \partial_t
\end {equation}
for the relevant degrees of freedom, which puts the effective equation
precisely in the form of (\ref{eq:molasses}).
The damping coefficient $\gamma$ is
\begin {equation}
   \gamma = \# {g^2 T^2 \over k}
   .
\label{eq:gamma}
\end {equation}

We can now see that the damping is strong.  Remember that topological
transitions are dominated by spatial momenta $k$ of order $g^2 T$.
Then (\ref{eq:gamma}) gives
\begin {equation}
   \gamma \sim T \gg k
   .
\end {equation}
The characteristic frequency of large-amplitude fluctuations is then
given by (\ref{eq:wc}):
\begin {equation}
   \omega_{\rm c} \sim {k^2\over\gamma} \sim g^2 T
   .
\end {equation}
The associated time scale is
\begin {equation}
   t \sim {1\over\omega_{\rm c}} \sim {1\over g^4 T}
   .
\end {equation}
This is the main result of this talk and is precisely the claim I
made at the beginning for the time scale relevant to non-perturbative
physics such as topological transitions.
Using the power spectrum (\ref{eq:power}), it is possible to check
(see ref.\ 2 for details)  
that the amplitude of fluctuations associated with $\omega_{\rm c}$ is
indeed non-perturbative---that is, that
\begin {equation}
   A \sim {1\over g L} \sim g T
   ,
\end {equation}
where the first inequality is the condition for non-perturbative
fluctuations and the second follows from $L \sim 1/g^2 T$.
Moreover, the fluctuations associated
with any larger frequency (shorter time scale) are smaller and
therefore perturbative---they cannot be responsible for topological
transitions.


\begin {figure}
\vbox
   {%
   \begin {center}
      \leavevmode
      \def\epsfsize #1#2{0.60#1}
      \epsfbox [150 300 500 530] {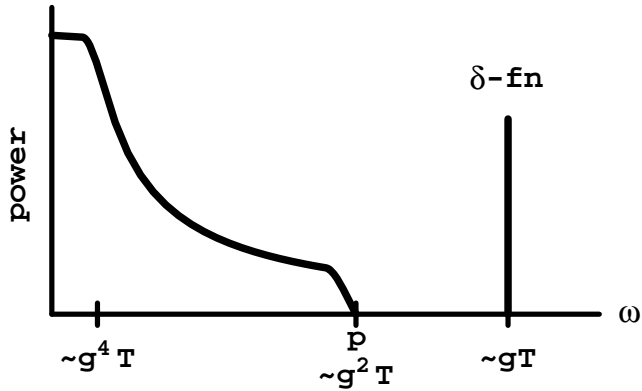}
   \end {center}
   \bigskip
   \caption
       {%
       A schematic plot of the power of gauge-field
       fluctuations with $k \sim g^2 T$.
       The $\delta$-function spike at $\omega \sim gT$ corresponds to
       propagating plasmons.
       \label{fig:color}
       }%
   }%
\end {figure}

\section {Putting the color back into white noise}

In the last section, I assumed that the relevant physics involved
$\omega \ll k$, which simplified the discussion by making the
effective equation of motion (\ref{eq:motion}) look just like the
idealized molasses example.  I now just want to show that nothing
drastic happens to the analysis if we don't assume $\omega \ll k$
from the beginning but keep the equation in the more general form of
(\ref{eq:motion}).  There is still a result from the fluctuation-dissipation
theorem, whose more general form looks like
\begin {equation}
   \langle \xi^* \xi \rangle \sim (n_\omega + 1) \, {\rm Im} \Pi \,
\end {equation}
where $n_\omega$ is the Bose distribution function.  The corresponding
response of $A$ may be written as
\begin {equation}
   \langle A^* A \rangle \sim (n_\omega+1) \; {\rm Im}
      \left( 1 \over (\omega+i\epsilon)^2 - k^2 - \Pi(\omega,k) \right)
   ,
\label{eq:color}
\end {equation}
where $\epsilon$ is an infinitesimal that selects the retarded response.
The resulting power spectrum is shown in fig.~\ref{fig:color}.
(See ref.\ 2 for details.)
For $\omega \ll k$, the result is the same as before; but the spectrum
disappears at $\omega = k$.
Then at $\omega \sim gT$ there is a
$\delta$-function corresponding
to the pole of (\ref{eq:color}), which corresponds to propagating plasma
waves.%
\footnote{
   This is all in the leading-order approximation to $\Pi$.  In reality,
   the disappearance of $\Im\Pi$ at $\omega=k$ will not be sharp, the
   $\delta$-function will have a width, and there will be two-plasmon cuts
   and other features to the spectrum.  All of these modifications,
   however, are sub-leading and do not affect the analysis being made
   here.
}
As before, one can check that the fluctuations in $A$ for all
frequencies with $\omega \gg g^4 T$ are only perturbative and cannot
be responsible for topological transitions.
Non-perturbative physics comes from $\omega \sim g^4 T$.


\section {Conclusions}

  I've tried to argue as pedagogically as possible that the time scale
associated with non-perturbative, colored fluctuations of hot gauge
fields is $t \sim 1/g^4 T$ and that the rate for baryon number violation
in the hot, symmetric phase of electroweak theory is therefore
$\Gamma \sim \alpha_w^5 T^4$ at weak coupling.  There are a couple of
interesting things I haven't have time to discuss:

\begin {itemize}

\item 
   I treated all the effective long-distance equations in linear
   approximation.  In fact, the full non-linear effective equations are
   known and come from an analysis of what are known as ``hard thermal
   loops.''  See ref.~1 for a review and Huet and Son,\cite{huet}
   as well as Son's talk in these proceedings.

\item
   This effective long-distance theory is {\it not} generally
   rotationally invariant
   for classical theories defined on a short-distance lattice, such as
   those discussed in the talks by Krasnitz and by Moore\,\cite{classical}
   for simulating
   the B violation rate numerically.  See ref.~1
   and Bodeker et al.\cite{bodeker}\ for a discussion.
   The failure to recover rotational invariance (which may come as a
   surprise to those readers more familiar with Eucliean-time lattice
   systems) turns out not to be a complete distaster for using classical
   lattice simulations to understand the continuum theory.\cite{arnold}

\item
   It is possible to convert between the rate $\Gamma$ measured in classical
   lattice simulations and the real $\Gamma$ in quantum field theory.
   See ref.~1.

\end {itemize}



\section*{Acknowledgments}
This work was supported by the U.S. Department of Energy, grant
DE-FG03-96ER40956.

\end{document}